\documentclass[twocolumn,secnumarabic,amssymb, nobibnotes,aps,prl,reprint]{revtex4}

\usepackage{graphicx}
\usepackage[dvips]{color}
\usepackage{amsmath,amsthm,amssymb}

\newcommand{\be}{\begin{equation}}
\newcommand{\ee}{\end{equation}}
\newcommand{\ba}{\begin{eqnarray}}
\newcommand{\ea}{\end{eqnarray}}

\renewcommand{\vec}[1]{\boldsymbol{#1}}

\def\lsim{\raise0.3ex\hbox{$\;<$\kern-0.75em\raise-1.1ex\hbox{$\sim\;$}}}
\def\gsim{\raise0.3ex\hbox{$\;>$\kern-0.75em\raise-1.1ex\hbox{$\sim\;$}}}

\def\theta{\vartheta}

\begin{document}

\title{Explaining the Spectra of Cosmic Ray Groups above the Knee by Escape from the Galaxy}

\author{G.~Giacinti$^{1}$}
\author{M.~Kachelrie\ss$^{2}$}
\author{D.~V.~Semikoz$^{3}$}
\affiliation{$^{1}$University of Oxford, Clarendon Laboratory, Oxford, United Kingdom}
\affiliation{$^2$Institutt for fysikk, NTNU, Trondheim, Norway}
\affiliation{$^3$AstroParticle and Cosmology (APC), Paris, France}

\begin{abstract}
We investigate the possibility that the cosmic ray (CR) knee is entirely explained by the
energy-dependent CR leakage from the Milky Way. We test this hypothesis
calculating the trajectories of individual CRs with energies between
$E/Z=10^{14}$\,eV and $10^{17}$\,eV propagating them in the regular and 
turbulent Galactic magnetic field.  
We find a knee-like structure of the CR escape time $\tau_{\rm esc}(E)$ 
around $E/Z={\rm few}\times 10^{15}$\,eV for a 
coherence length $l_{\rm c} \simeq 2$\,pc of the turbulent field, 
while the decrease of $\tau_{\rm esc}(E)$ slows down around 
$E/Z\simeq 10^{16}$\,eV in models with a weak turbulent magnetic field. 
Assuming that the injection spectra of CR nuclei are power-laws, the 
resulting CR intensities in such a turbulence are consistent with the energy spectra of CR nuclei 
determined by KASCADE and KASCADE-Grande. 
We calculate the resulting CR dipole anisotropy as well as the source
rate in this model.
\end{abstract}


\maketitle



\textit{Introduction.}---The cosmic ray (CR) energy spectrum follows 
a power-law on more than ten decades in energy. Only a few breaks, 
such as the knee at $E_{\rm k} \approx 4$\,PeV, provide possible clues
to how CRs propagate and to what their sources are. In addition 
to this feature, seen
in the total CR flux, the elemental composition of the CR flux
in the energy range $10^{15}$--$10^{17}$\,eV is especially useful to 
constrain theoretical models for the knee. Such data had been missing,
but recently the KASCADE-Grande collaboration has extended the measurements 
of the intensity of individual groups of CR nuclei up to $10^{17}$\,eV 
\cite{KG,dataKG}. In the future, the IceCube experiment with its IceTop 
extension
will provide additional constraints on the mass composition 
around the knee~\cite{top}. It is therefore timely to
compare not only the  predictions for the total CR flux and global 
quantities such as the elongation rate to the experimental data, 
but also to consider the intensity of individual groups of CR nuclei.

Three competing
explanations for the origin of the knee have been advanced: First, there 
have been speculations that interactions may change in the multi-TeV 
region, 
thereby suppressing
the CR flux. This possibility is now excluded by the LHC data.
Second, the knee may correspond to the maximum rigidity to which 
the dominant population of Galactic CR sources can accelerate 
CRs~\cite{pop}. In a variation of this suggestion, the knee
is caused by the maximal energy of a single nearby source such as 
Monogem~\cite{Erl}. While it is very natural to expect
that differences in supernovae types and their environments lead
to a distribution of reachable maximal rigidities, this proposal does not 
predict the exact energy of the knee or the strength of the flux 
suppression, without a better knowledge of 
CR confinement and escape~\cite{Bell:2013kq}.

Finally,  the knee may be caused by a change of the diffusion properties
of charged CRs~\cite{sa,hall}. For instance, the knee may correspond to 
the rigidity at 
which the CR Larmor radius $r_{\rm L}$ starts to be of the order of the 
coherence length $l_{\rm c}$ of the turbulent Galactic magnetic field (GMF). 
Thence the behaviors of the CR diffusion coefficient and confinement time 
change, which in turn would induce a steepening in the spectrum. 
Both small-angle scattering~\cite{sa} and Hall diffusion~\cite{hall} have 
been proposed as models for the energy dependence of the diffusion coefficient 
in this regime. This phenomenological approach to CR escape that describes
CR propagation by diffusion has two major drawbacks: 
First, the analytical connection between the diffusion tensor
and the underlying magnetic field is know only in certain limiting regimes. 
Second, the diffusion approximation is not justified at the highest 
energies we are interested in.

The goal of this Letter is therefore to study CR escape from our Galaxy 
by propagating individual CRs in detailed GMF models. Given a specific 
GMF model, this approach allows us to predict the 
position and the shape of the knee for CR nuclei as a function of $E/Z$.
We determine three main observables: 
The time-averaged grammage $X(E)$ traversed by CRs before escape, the 
time-dependent intensity of CR nuclei at the position of the Sun, and
the amplitude $d$ of the dipole anisotropy. Extrapolating $X(E)$ 
towards its measured value at low energies, we constrain first the
strength of the turbulent Galactic magnetic field. Then we use the 
CR dipole anisotropy as an indication for the transition from Galactic 
to extragalactic CRs.  Finally, we calculate the CR 
nuclei energy spectra, and compare them to the intensity of 
groups of CR nuclei determined by KASCADE and KASCADE-Grande, 
examining the consistency of the proposed scenario.
Our results suggest that the turbulent GMF is weaker than in
GMF models as~\cite{J} and has a small coherence length. Improving
our undestanding of the GMF has important impacts on fields outside
CR research as  e.g.\ indirect DM searches, where our results
support the use of a large GMF halo and a large diffusion coefficient.

{\it Simulation procedure---}%
For the propagation of CRs in the GMF, we use the code described and tested 
in~\cite{Giacinti:2011ww}. For the present work, we have implemented the
most recent GMF models~\cite{P,J},  but a more important change is the 
use of a reduced coherence length, $l_{\rm c}=2$\,pc, for the turbulent 
field. Such a value is in line with recent measurements of $l_{\rm c}$ in 
some regions of the Galactic disk, see e.g.~\cite{Iacobelli:2013fqa}. 
We model the random
field as isotropic Kolmogorov turbulence, $\mathcal{P}(k)\propto k^{-5/3}$ 
extending down to a minimal length scale $l_{\min} \lesssim 1$\,AU.

We assume that the surface density of CR sources follows the distribution
of supernova remnants~\cite{Green:2013qta},
\be  \label{SNdist}
 n(r) = \left( r/R_\odot \right)^{0.7} 
        \exp\left[-3.5(r-R_\odot)/R_\odot\right] \,,
\ee
with $R_\odot=8.5$\,kpc as the distance of the Sun to the Galactic center.

{\it Grammage}---%
An important constraint on CR propagation models comes from ratios of 
stable primaries and secondaries produced by CR interactions on gas 
in the Galactic disk. In particular, the B/C ratio has been recently 
measured by the AMS-02 experiment up to 670\,GeV/nucleon~\citep{AMS02}. 
Above $E \gtrsim 10$\,GeV, it is consistent with a straight power-law.

Using the leaky-box formalism, various propagation models including plain 
diffusion, diffusion with convection and reacceleration were fitted in
Ref.~\cite{Jones} to earlier data. In all cases, the grammage traversed 
by CRs at 
reference energies $E_0/Z=5$--15\,GeV was found to lie in the range 
9--14\,g/cm$^2$. In their model with reacceleration, the grammage decreases 
as $X(E)=X_0 (E/E_0)^{-0.3}$ above energies $E_0>20$--30\,GeV per nucleon, 
while the best-fit value for the normalization constant $X_0$ of the 
grammage was determined to be $X_0=9.4$\,g/cm$^2$.

We calculate the average grammage 
$\langle X\rangle =N^{-1}c\sum_{i=1}^N\int dt\,\rho(\vec x_i(t))$ 
by injecting $N$ cosmic rays according to the radial distribution 
(\ref{SNdist}) 
at $z=0$ in the Galaxy, and  following their trajectories $\vec x_i(t)$ 
until they reach the edge of the Galaxy. 
As a model for the gas distribution in the Galactic disk, we use 
$n(z)=n_0\exp(-(z/z_{1/2})^2)$ with $n_0=0.3/$cm$^3$ at $R_\odot$ 
and $z_{1/2}= 0.21$\,kpc, inspired by \cite{gas}. 
We set $n=10^{-4}$g/cm$^3$ as minimum gas density up to the edge of 
the Milky Way at $|z|=10$\,kpc. Since the grammage 
$X(E)\propto E^{-\delta}$ scales as the confinement time 
$\tau(E)\propto E^{-\delta}$, we can use this quantity also as an
indicator for changes in the CR intensity induced by a variation
of the CR leakage rate.

\begin{figure}
  \includegraphics[width=0.45\textwidth,angle=0]{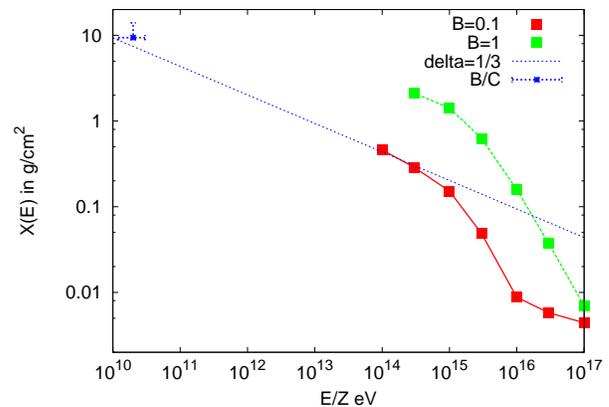}
  \caption{Grammage $X(E)$ traversed by CR protons as a function of energy $E/Z$ 
    for two different levels of magnetic turbulence in the 
    GMF model of Ref.~\cite{J}, with $l_{\rm c}=2$\,pc.}
  \label{figI}
\end{figure}

In Fig.~\ref{figI}, we show the grammage traversed by CRs, with energies 
$E/Z$ between $10^{14}$\,eV and $10^{17}$\,eV, propagated in 
the GMF model of Ref.~\cite{J}. 
The upper (green) line corresponds to computations using both 
the regular and the turbulent fields proposed in~\cite{J}, 
while for the lower (red) curve we rescaled 
the turbulent field strength by a factor 0.1 ($B_{\rm rms}\to B_{\rm rms}/10$). 
The two lowest-energy data points shown here are consistent with the 
$X(E)\propto E^{-1/3}$ behavior expected for a turbulent magnetic field 
with a Kolmogorov power-spectrum. 
Around a few PeV, the grammage steepens to an approximate 
power-law $X(E)\propto E^{-1.3}$ which lies in-between the expectations 
for Hall diffusion ($X(E)\propto E^{-1}$) and small-angle scattering 
($X(E)\propto E^{-2}$). 
The transition energy  agrees well with the theoretical expectation: 
A steepening of the grammage is
expected at the characteristic energy $E_c$ where the Larmor radius 
$r_{\rm L}$ equals the coherence length $l_c$. For $l_c=2$\,pc and 
$B\approx 3\,\mu$G, the value of the  critical energy is 
$E_c/Z\approx 6\times 10^{15}$\,eV, which is only slightly higher than 
what we find numerically. 
Finally, the turnover of the grammage which is visible in the lower curve 
at the highest energies corresponds to the approach of its asymptotic value 
obtained for straight line  propagation in the limit $E\to\infty$.
As a consequence, the predicted CR spectra above $E/Z\simeq 10^{16}$\,eV should 
harden by approximately one power, $\Delta\delta\sim 1$, using the
model of Ref.~\cite{J} with a {\em reduced\/} turbulent
field. 

In addition to the data points, we show in Fig.~\ref{figI} with a 
dotted (blue) line the extrapolation 
of the grammage to lower energies, assuming that 
$X(E)\propto E^{-1/3}$ as expected for a Kolmogorov power-spectrum. 
Based on this extrapolation, the GMF model~\cite{J}  with full turbulence 
leads to a grammage at $E \lesssim 100$\,GeV which is a factor $\sim 10$ above 
the determinations from e.g.\ B/C measurements (blue cross). 
This discrepancy is in line with our determination of the diffusion 
coefficient in a purely turbulent magnetic field with strength 
$B_{\rm rms}=4\,\mu$G \citep{Giacinti:2012ar}, which also disagreed by 
an order of magnitude with the extrapolation of the diffusion coefficient 
phenomenologically determined from the ratios of secondary to primary nuclei.
Consistency with these 
measurements could be achieved, if the energy density of the turbulent 
magnetic field is reduced~\footnote{The GMF model of Ref.~\cite{J} is 
currently being 
revised, taking into account correlations between the electron density and 
the magnetic field. As a result, the turbulent field will be reduced
by a factor 5--6 (G.~Farrar, private communication).}. Such a rescaling 
is displayed as the lower (red) line in Fig.~\ref{figI}. 
In the following,
we consider this case~\footnote{We have also performed calculations
for $l_c=5$\,pc. In this case, the turbulent field has to be rescaled only
by a factor 0.2.}.

We have calculated the grammage also in the GMF model 
of Ref.~\cite{P}. The CR confinement time 
was found to be twice as large as for the GMF model 
of Ref.~\cite{J}, leading 
to a stronger discrepancy between the extrapolation of the grammage 
to low energies and its determinations.

\begin{figure}
  \includegraphics[width=0.47\textwidth,angle=0]{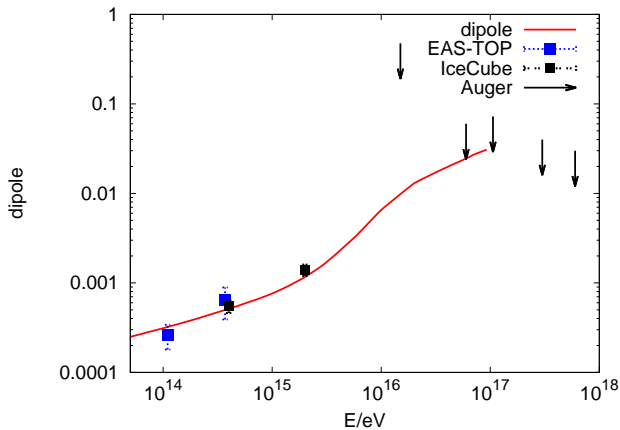}
  \caption{Dipole amplitude $d(E)$ as a function of energy 
    $E/Z$ in the GMF model of Ref.~\cite{J} with
    reduced turbulent field and $l_{\rm c}=2$\,pc.}
  \label{figd}
\end{figure}

{\it Cosmic ray anisotropy---}%
In the diffusion approximation, the CR dipole anisotropy $d$ is given by 
$\vec d = 3D \: \vec\nabla\ln(n) /c$. 
The measurements or tight experimental upper limits on $d$ at high energies are typically 
difficult to reconcile with determinations of the diffusion coefficient 
at low energies, even for a Kolmogorov spectrum where $D(E) \propto E^{1/3}$. 
In our model, one expects the anisotropy to grow more rapidly above the knee. 
The diffusion coefficient scales there as $D \sim \frac13 l_{\rm c} c (R_{\rm L}/l_{\rm c})^\alpha$, 
with $\alpha \approx 1.3$. We compute the average anisotropy and derive 
the energy dependence of $D(E)$ from the escape probability calculated 
previously, setting $D(E/Z)\propto\tau_{\rm esc}(E/Z)$. We fix the 
proportionality constant by requiring that the dipole amplitude 
$d=\sum_k f_k d_k$ equals the dipole component $\tilde d$ observed by the EAS-TOP 
collaboration at $E=1.1\times 10^{14}$\,eV~\cite{EASTOPD}. $k$ labels the groups 
of nuclei we consider, $f_k$ is their fraction of the total CR flux, 
and $d_k\propto \tau_{\rm esc}(E/Z)$ is their individual dipole.

In Fig.~\ref{figd} we show the resulting dipole amplitude $d$ as a 
function of energy $E$. As expected, the amplitude raises below the knee as $E^{1/3}$, 
while it increases approximately as $E^{0.7}$ above. 
We also plot the values of $\tilde d$ observed by 
IceCube~\cite{Ianisp}, as well as the 99\% C.L. upper limits on $d_\perp$ from the Pierre Auger 
Observatory~\cite{PAOaniso}. Comparing our estimate 
for the dipole amplitude with the upper limits 
at high energies, we conclude that the light component of the 
Galactic CR flux must be suppressed above $10^{17}$\,eV.
We expect the approximation $d\propto \tau_{\rm esc}(E/Z)$ to 
break down for $E/Z \gtrsim 10^{17}$\,eV : Close to the semi-ballistic regime, a calculation of 
the anisotropy based on trajectories would be required, but is computationally extremely expensive.

{\it Cosmic ray intensity---}%
We now use the time-dependent intensity $I_{k}$ of various groups $k$ of CR
nuclei as a test of our hypothesis that the knee is entirely 
explained by the energy-dependent CR leakage from the Milky Way. 
We distribute a discrete set of sources in the Galactic disk according to 
Eq.~(\ref{SNdist}) and a fixed rate $\dot N$. Each source is assumed
to inject the total energy $E_{\rm p} = 1.0 \times 10^{50}$\,erg in CRs. 
Then the individual contributions 
$n_{i,k}(\vec x,t,E)$ from each source $i$ are added using a pre-calculated 
template in order to save computing time. We convert the total 
density $n_{k}(\vec x_\odot,t,E)$ at the position of the Sun into 
the predicted intensity 
$I_{k}(\vec x_\odot,t,E)=c/(4\pi)n_{k}(\vec x_\odot,t,E)$ of the CR nuclei
group $k$ as a function of time. Finally,
we determine the relative fraction of energy transferred to the $k$.th 
group of nuclei and the exponent $\alpha_k$ of their injection spectrum 
$dN_k/dE\propto E^{-\alpha_k}\exp(-E/ZE_c)$ by a comparison of the predicted 
intensity to the measurements. We choose the energy $E_c$ above which we
assume that the source spectrum is exponentially suppressed as
$E/Z=10^{17}$\,eV.

Note that $I_k(E,t)$ is predicted as a function of time. Since $I_k(E,t)$ 
fluctuates due to the discreteness of the sources, we show 
a $1\sigma$ confidence band illustrating the spread around the predicted
average intensity. In contrast, the 
grammage is measured at relatively low energies, $E \lesssim 1$\,TeV, where 
fluctuations due to the discrete source distribution play only a minor role
and therefore only average values are relevant.

For the experimental data, we use above $E>10^{16}$\,eV the 
tabulated intensities of protons, helium, carbon (representing the CNO group),
Si (for the SiMg group) and Fe nuclei (for Fe-Mg) from KASCADE-Grande data 
given in~\cite{dataKG}, while we employ in the energy range  
$E=10^{15}-10^{16}$\,eV  the KASCADE data read from Fig.~12 in~\cite{dataKG}.
Since the KASCADE and KASCADE-Grande proton data disagree in the
overlapping energy range, we reduce the KASCADE-Grande proton flux by 30\%, and 
add this 30\% proton flux to the He flux, 
in order to achieve agreement between the two data sets. 
Moreover, the discriminating power between Si and Fe in KASCADE 
was relatively poor and the absolute Si+Fe abundance is small close to 
$10^{15}$\,eV, and therefore we analyze these two groups of elements
jointly~\cite{AH,Antoni:2005wq}. Finally, we determine the slope $\alpha_k$ and the
normalisation of the intensities for individual CR groups with
measurements at energies below the knee using the data from the CREAM 
experiment~\cite{CREAM}.

\begin{figure*}
\includegraphics[width=0.34\textwidth,angle=270]{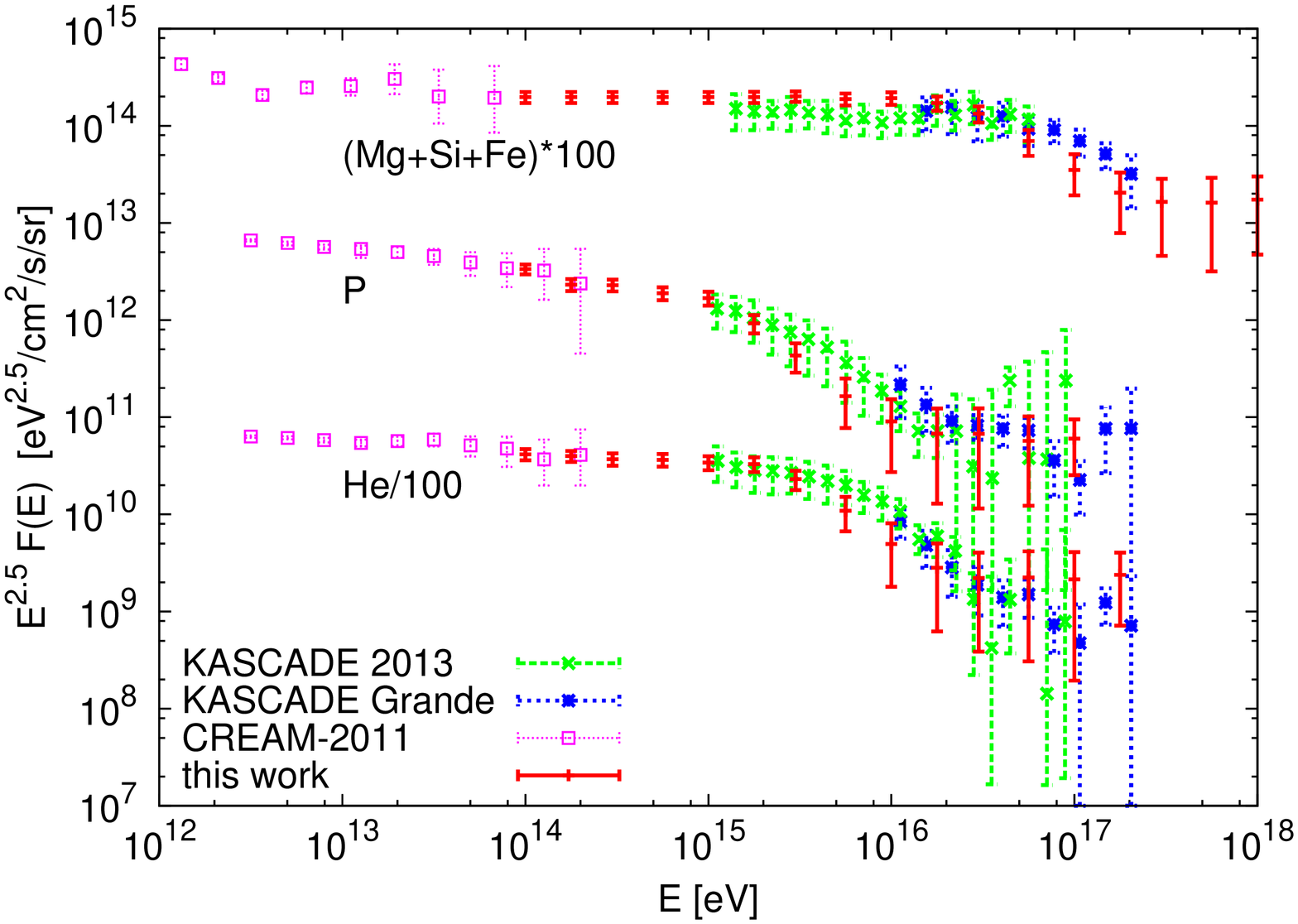}
\hskip0.4cm
\includegraphics[width=0.34\textwidth,angle=270]{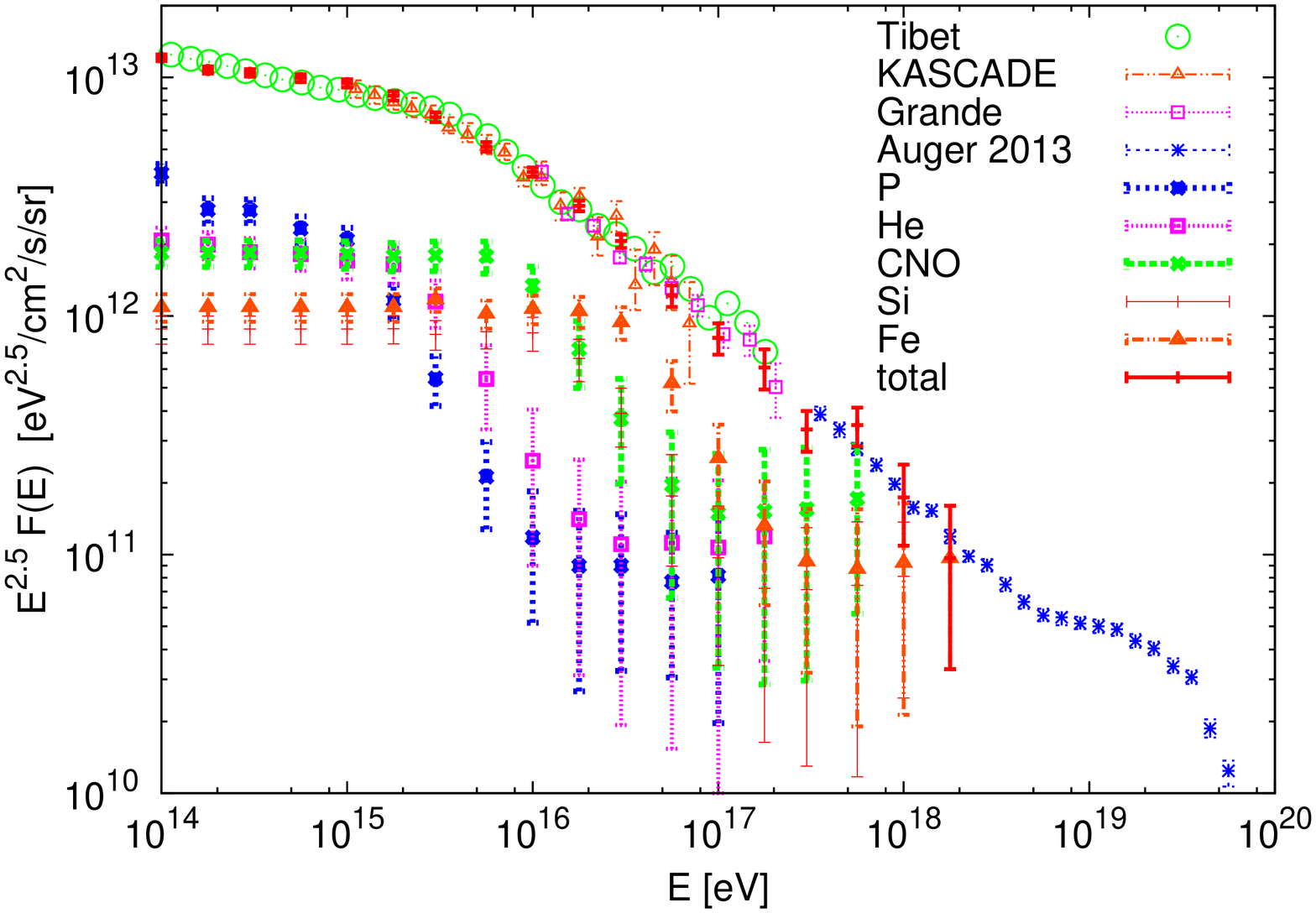}
\caption{Left: 
The (rescaled) intensity $I(E)$ of CR protons, He and 
Fe nuclei compared to the experimental
data from KASCADE~\cite{dataKG}, KASCADE-Grande~\cite{dataKG} and
CREAM~\cite{CREAM}, using the rescaled turbulent GMF. 
Right: Intensity $I(E)$ of CR protons, He, CNO, MgSi and Fe nuclei
as a function of energy $E$ per nucleus, obtained using the same GMF.}
\label{figII}
\end{figure*}

In the left panel of Fig.~\ref{figII}, we show our results (red solid points with 
error-bars) for the intensity of p, He, and the combined intensity of the 
MgSi and Fe groups compared to experimental data. Overall, we find
that the assumptions used, a GMF model with rescaled Kolmogorov turbulence 
and $1/E^{\alpha_k}$ injection spectra, lead to a consistent CR intensity
in the full energy range $E\sim (10^{14}-10^{17})$\,eV considered not only for 
protons, but also for He and heavy nuclei. Since CR escape depends only 
on rigidity, and since the exponents $\alpha_k$ are determined by the data 
below the knee, the relative shape of the different CR elements is fixed
in this scheme. Note also that the recovery of the proton and helium spectra
above $E/Z\sim 10^{16}$\,eV cannot be reproduced assuming power-law 
injection spectra and the full turbulent GMF of Ref.~\cite{J}.

In the right panel of Fig.~\ref{figII}, 
we show the contribution of the CNO group and the 
resulting total CR intensity. The latter is compared to measurements of the total CR intensity by 
Tibet~\cite{Amenomori:2008aa}, KASCADE, KASCADE-Grande and Auger~\cite{Aab:2013ika}: Because of the
rigidity-dependent energy cutoff, Galactic CRs are dominated
at the highest energies by iron, while the total intensity is
exponentially suppressed above $3\times 10^{18}$\,eV. Thus 
Galactic Fe could give a significant contribution to 
the total CR spectrum up to the ankle, being especially 
important for composition studies.  

The obtained source rate $\dot N\sim 1/180$\,yr is only a factor six 
smaller than the Galactic SN rate and makes a GRB origin of Galactic 
CRs unlikely. Taken at face value, these numbers would require that 
a large fraction of SNe can accelerate protons up to $E\sim 10^{17}$\,eV.

{\it Conclusions---}%
The two main explanations for the knee are $i)$ a change in the CR 
confinement time 
in the Galaxy when their Larmor radius starts to be of the order of 
the coherence length $l_{\rm c}$ of the interstellar turbulence, and 
$ii)$ a change in the number of 
sources able to accelerate CRs above $\approx 4$\,PeV. 
We have shown that, if the coherence length 
in the Galactic disk is of the order of $l_{\rm c}=2$\,pc as suggested 
by Refs.~\cite{Iacobelli:2013fqa}, the CR escape time 
$\tau_{\rm esc}(E)\sim X(E)$ 
and as a consequence the total CR intensity steepens at the correct energy.
Moreover, the resulting rigidity-dependent knees in the
individual intensities of the considered CR groups (p, He, CNO, and MgSiFe) 
agree well with measurements. 

In contrast, the change in the slope of $X(E/Z)\sim \tau(E/Z)$ would 
be shifted to energies above $E/Z\simeq 10^{16}$\,eV for a coherence length 
$l_{\rm c}$ of the order of $l_{\rm c}=50$\,pc in the Galactic disk.
In this case, the knee would have to be explained by the possibility $ii)$, 
yielding precious information on Galactic CR accelerators. More measurements 
of the coherence length of the turbulence as expected from e.g.\ 
SKA~\cite{SKA} will solve this crucial question.

A large coherence length $l_{\rm c}$ would worsen the tension 
between our computations of the grammage in GMF models like \cite{J,P} and
its determination from B/C at low energies. 
In this Letter, we have therefore considered the possibility 
that the average strength of the turbulent magnetic field is reduced 
by a factor $\simeq 5-10$. In this case, we found agreement between
our calculation of the escape time $\tau_{\rm esc}$ and determinations 
of the grammage $X(E)$ at lower energies from the B/C ratio. More
importantly, the turnover of the grammage $X(E)$ leads to 
a hardening of the intensity of the nuclei with charge $Z$ around 
$E/Z\simeq 10^{16}$\,eV. This energy behavior is very similar to the
one seen in KASCADE-Grande measurements~\cite{KG}: As a result, the observed energy
dependence of various groups of CR nuclei can be explained in the energy
range between $10^{14}$ and $10^{17}$\,eV as the modulation of power-law
injection spectrum via CR leakage from the Galaxy.

The source rate in our scenario is relatively large, $\dot N\sim 1/180$\,yr,
and requires that a large fraction of SNe can accelerate protons up to 
$E\sim 10^{17}$\,eV. Our estimate for the dipole anisotropy 
suggests on the other hand that CRs with energies beyond 
$E/Z\sim 10^{17}$\,eV are not predominantly Galactic, 
which requires an early transition to extragalactic CRs, before 
the ankle as e.g.\ in the dip model~\cite{dip}. A hint about this 
is also given by KASCADE-Grande~\cite{Apel:2013ura}. 
Finally, we note that the suggested weakness of the turbulent GMF 
would facilitate the search for ultra-high energy CR sources by a smaller 
spread of their images on the sky~\cite{Giacinti:2011uj}.

\acknowledgements

We thank P.~Blasi for raising the grammage question, and
S.~Ostapchenko and A.~Strong for useful discussions. We are grateful 
to A.~Haungs for advice about the KASCADE data.
GG acknowledges funding from the  European Research Council 
(Grant agreement No. 247039). 
The work of DS was supported in part by grant RFBR \# 13-02-12175-ofi-m.



\begin{thebibliography}{00}
\bibitem{EP}
  Note that the experimental values $\tilde d$ are connected to the true dipole amplitude $d$ as $\tilde d= d_\perp/\langle\cos(\delta)\rangle$, where $d_\perp$ is the dipole component in the equatorial plane and $\langle\cos(\delta)\rangle$ the average declination of the events used in the harmonic analysis based only on R.A.; See 
  J.~Aublin and E.~Parizot,
  Astron.\ Astrophys.\ {\bf 441}, 407 (2005).
\end{thebibliography}
\end{document}